\documentclass[fleqn,usenatbib]{mnras}

\usepackage{newtxtext,newtxmath}

\usepackage[T1]{fontenc}

\usepackage{graphicx}	
\usepackage{amsmath}	

\newcommand{\imw}{$i$--$W3$}

\newcommand{\imwt}{$i$--$W2$}

\newcommand{\kms}{km~s$^{-1}$}

\newcommand{\kmsMpc}{km~s$^{-1}$~Mpc$^{-1}$}

\newcommand{\lam}{$\lambda$}
\newcommand{\Lam}{$\Lambda$}
\newcommand{\Om}{$\Omega$}

\newcommand{\hb}{H$\beta$}

\newcommand{\lya}{\mbox{Ly$\alpha$}}

\newcommand{\heii}{\mbox{He\,{\sc ii}}}

\newcommand{\civ}{\mbox{C\,{\sc iv}}}

\newcommand{\oiii}{\mbox{O\,{\sc iii}}}

\newcommand{\oi}{\mbox{O\,{\sc i}}}

\newcommand{\feii}{\mbox{Fe\,{\sc ii}}}

\newcommand{\mgii}{\mbox{Mg\,{\sc ii}}}

\title[ERQ Redshifts \& Outflows]{Accurate Systemic Redshifts and Outflow Speeds for Extremely Red Quasars (ERQs)}

\author[J. Gillette et al.]{
Jarred Gillette,$^{1}$\thanks{E-mail: jgill016@ucr.edu }
Fred Hamann,$^{1}$
Marie Wingyee Lau,$^{1}$
Serena Perrotta$^2$
\\
$^{1}$Department of Physics \& Astronomy, University of California, Riverside, CA 92521, USA\\
$^{2}$Center for Astrophysics and Space Sciences, University of California, San Diego, CA 92093, 
USA\\
}

\date{Accepted XXX. Received YYY; in original form ZZZ}

\pubyear{2023}

\begin{document}
\label{firstpage}
\pagerange{\pageref{firstpage}--\pageref{lastpage}}
\maketitle

\begin{abstract}
Extremely Red Quasars (ERQs) are thought to represent a brief episode of young quasar and galactic evolution characterized by rapid outflows and obscured growth due to dusty environments. We use new redshift measurements from CO and \lya\ emission-lines to better constrain outflow velocities from previous line measurements. We present sample of 82 ERQs, and the analysis confirms that ERQs have a higher incidence of large \civ\ blueshifts, accompanied by large Rest Equivalent Widths (REWs) and smaller line widths than blue quasars. We find that strong blueshifts (>2000 \kms) are present in 12/54 (22.22\%) of ERQs with the most robust redshift indicators. At least 4 out of 15 ERQs in the sample also have blueshifts in their \hb\ and low-ionization UV lines ranging from $-$500 to $-$1500 \kms. ERQs with strong \civ\ blueshifts are substantially offset in \civ\ REW and Full-Width at Half-Maximum (FWHM) from typical blue quasars in the same velocity range. ERQs have average values of REW~=~124\AA\ and FWHM~=~5274 \kms, while blue quasars have REW~=~24\AA\ and FWHM~=~6973 \kms. The extreme nature of the outflows in ERQs might explain some of their other spectral properties, such as the large \civ\ REWs and peculiar wingless profiles owing to more extended broad-line regions participating in outflows. The physical reasons for the extreme outflow properties of ERQs are unclear; however, larger Eddington ratios and/or softer ionizing spectra incident on the outflow gas cannot be ruled out.

\end{abstract}




\section{Introduction}

Quasars are supermassive black holes rapidly growing by accretion of infalling material at the center of their host galaxy. Accretion can coincide with streams of infalling gas, or galactic assembly, at high redshift. Merger activity, or infalling gas, may be the triggers supplying matter for starbursts and quasar activity \citep{Hopkins+06,Hopkins+08,Somerville+08,Keres+09,Dekel+09,FaucherGiguereKeres11,Fumagalli+14,Glikman+15}. This rapid accretion can coincide with outflows, generating feedback, and potentially influence the galaxy's formation \citep{CostaSijackiHaehnelt14,Nelson+15,Suresh+19}. A possible evolutionary scheme is where the central black hole initially grows in obscurity until feedback generates outflows, and clears the obscuring interstellar material, revealing a normal blue quasar \citep{Sanders+88,DiMatteo+05, Hopkins+06,Hopkins+08,Hopkins+16,RupkeVeilleux11,RupkeVeilleux13,Liu+13,Stacey+22}.

Red quasars are important for testing the hypothesis that obscured quasars may be in a young, and short-lived, phase in their evolution. Young quasars may be reddened by dust created in a major starburst inside the galaxies, triggered by a merger or cold-mode accretion. These reddened quasars may show signs of youth such as high accretion rates, more infall from the intergalactic medium, or powerful outflows, as they transform in to blue/unobscured quasars. Studies of red/obscured quasars have found many to be in mergers or high-accretion phases \citep{Glikman+15,Wu+18,Zakamska+19}. 

Extremely red quasars (ERQs) are a unique quasar population selected from the Baryon Oscillation Spectroscopic Survey \citep[BOSS,][]{Paris+17} in the Sloan Digital Sky Survey-III \citep[SDSS,][]{Eisenstein+11} and the ALLWISE data release \citep{Cutri+11,Cutri+13} of the Wide-field Infrared Survey Explorer \citep[WISE,][]{Wright+10}, based on red rest-UV to mid-IR colors, \imw $>$ 4.6 \citep[AB magnitudes][]{Ross+15, Hamann+17}. They are of particular interest because of a suite of other spectral properties connected to their extreme red colors, namely, unusually strong \civ\ \lam 1549 emission lines with peculiar wingless profiles and frequent large blueshift, a high incidence of broad \civ\ outflow absorption lines, and exceptionally fast outflows measured in [\oiii] \lam 4959,5007 \citep{Hamann+17,Perrotta+19}. All of these features might be explained by exceptionally powerful accretion-disk outflows, e.g., that produce spatially-extended \civ\ broad emission line regions that produce large \civ\ rest equivalent widths (REWs) and feed into the fast, lower-density [\oiii] outflows farther out \citep{Zakamska+16, Hamann+17, Perrotta+19}. This evidence for prodigious outflows, combined with the extreme dust reddening, make ERQs prime candidates for quasars driving feedback to their host galaxies during the early stages of massive galaxy evolution. 

An essential ingredient for understanding quasar outflows, and the exotic properties of ERQs in particular, is accurate systemic redshifts to define the outflow speeds and kinetic energies. The most accurate redshift indicators are narrow emission lines, like [\oiii] \lam 5007, that form in the extended galactic or circumgalactic environments of quasars. When these lines are not available, \hb\ and various low-ionization broad emission lines in the UV, such as \mgii\ \lam 2800, \oi\ \lam 1304, can be useful for normal quasars because they are less likely to be blueshifted (i.e., participating in outflows) than high-ionization emission lines like \civ\ \citep[e.g.][]{Shen+08,Shen+16,Li+17}.

However, these redshift indicators are often problematic for ERQs because outflow signatures are more common and more extreme throughout their spectra. In particular, their [\oiii] lines are typically very broad and blueshifted with no distinct narrow components that might form in the extended/galactic environments \citep{Zakamska+16, Perrotta+19}, and their low-ionization broad emission lines often have large measurement uncertainties due noisy  spectra and they might be affected by large blueshifts/outflow speeds throughout the broad emission-line region (including more than just the high-ionization gas, see Section \ref{sec:sec_emission_speeds} below). 

In this paper, we provide reliable systemic redshifts for a sample of 82 ERQs that have existing measurements of CO in their host galaxies (Hamann et al. 2023 in preparation) and/or narrow \lya\ emission lines \citep[][and this work]{Lau+22,Gillette+23a} that form in their extended circumgalactic and/or halo environments. We then use these data to derive blueshifts/outflow speeds for the ERQ \civ\ \lam1549 broad emission lines and, whenever possible, reassess the [\oiii] \lam5007 outflow speeds reported previously by \citep[][and by Lau~et~al. in prep.]{Zakamska+16, Perrotta+19}. Our findings provide more accurate results, but support previous claims that ERQs have an unusually high frequency of fast/powerful outflows in these emission line regions, e.g., compared to normal blue quasars at similar redshifts and luminosities.

This paper is organized as follows. Section \ref{sec:sec_sample} describes selection criteria for quasars from catalog spectra and their redshifts. Section \ref{sec:sec_emission_speeds} describes our revised \civ\ and [\oiii] emission-line blueshifts and outflows. Section \ref{sec:sec_discuss} discusses and summarizes correlations in measured parameters and emission features, such as blueshift, emission strength, and line width. Throughout this paper we adopt a \Lam-CDM cosmology with $H_0$ = 69.6 \kmsMpc, \Om$_{\text{M}}$ = 0.286 and \Om$_\Lambda$ = 0.714, as adopted by the online cosmology calculator developed by \citet{Wright06}. All magnitudes are on the AB system. Reported wavelengths are in vacuum and in the heliocentric frame. 

\section{Sample Selection \& Systemic Redshifts}
\label{sec:sec_sample}

The parent sample for our study is the 205 ERQs selected to have \imw~$> 4.6$ in the combined BOSS survey \citet{Hamann+17}. Those authors provide detailed data for the \civ\ emission lines, including velocity shifts relative to the BOSS DR12 catalogue redshifts (see their Appendix A). The DR12 redshifts derive from automated fits to the BOSS spectra, plus algorithmic corrections based on nominal line blueshifts, that can be off by hundreds or $>$1000 \kms\ for quasars like ERQs with unusual emission-line properties. 

For the present study, we select ERQs from \cite{Hamann+17} that have other, more reliable measurements of the quasar systemic redshifts. These include 14 ERQs with ALMA sub-mm CO(4-3) emission-line measurements by Hamann et al. (2023 in prep.), 6 ERQs with KCWI (integrated field spectroscopy) of the extended \lya\ halos by Gillette et al. (submitted), and 54 ERQs with narrow \lya\ emission spikes in their BOSS spectra caused by halo emission. For some of our analysis, we also include 8 ERQs that have well-measured \mgii\ emission lines in their BOSS spectra, as measured by Gillette~et~al. (2023c in prep.).

Table \ref{tab:tab_catalogue} lists some basic catalogue properties and measurements for the total sample of 82 ERQs. The total ERQ sample has median color \imw\ $\approx$ 5.3 and redshifts in the range $1.8 \le z \le 3.7$. The subsample of 20 ERQs with $z_{\rm best}$ from either CO or Ly$\alpha$-halo has median color \imw\ $\approx$ 5.7, while the 54 ERQs with only a \lya-spike redshift have median color \imw\ $\approx$ 5.1. The small color difference between these two subsamples can be attributed to two factors beyond counting statistics in the samples sizes: 1) our preference for redder ERQs in the ALMA CO and KCWI \lya-halo observations, and 2) redder ERQs having fainter rest-UV fluxes that lead to noisier BOSS spectra and greater difficulty in identifying a narrow \lya-spike for our study (Section \ref{sec:sec_sample}). 

We include a large comparison sample of 39,909 (mostly) blue quasars from the SDSS survey with systemic redshifts measured from the \mgii\ \lam2800 broad emission lines (Gillette~et~al. 2023c in prep.). \mgii\ provides reasonably accurate systemic redshifts for normal blue quasars. For example, \cite{Shen+08} report that, in a large sample of SDSS quasars, the \mgii\ line has average blueshift relative to [\oiii] of $-$97 \kms\ and a dispersion in the measured shifts of 269 \kms. The sample with \mgii\ measurements are from emission-line profile fitting to SDSS BOSS spectra, is done with similar methodology as described in \citet{Hamann+17}, and with care taken to exclude bad data with custom signal-to-noise measurements and rejection criteria. \mgii\ emission is fit with a symmetric Gaussian or double-Gaussian profile, after the \feii\ complex near \mgii\ is removed using a \feii\ emission template. Details and results of the fitting in the blue quasar sample to will be presented in Gillette~et~al. (2023c in prep). \civ\ blueshifts are determined for the blue quasar sample by calculating the velocity shift of its centroid from the \mgii\ profile centroid.

\begin{table*}
\begin{center}
\caption{ERQ Sample Properties. $z_{em}$ is from the SDSS DR12Q BOSS catalogue emission-line measurement. $W3$ magnitude, and \imwt color are from WISE and BOSS. $z_{\text{best}}$ is the best estimate of the reference systemic redshift based on either CO (from Hamann~et~al. 2023 in prep.), narrow \lya\ emission lines in the extended ERQ halos \citep[from][]{Gillette+23a}, or narrow \lya\ emission-line ``spikes'' as measured here (Section 2) from the BOSS spectra. \civ\ REW, Full-Width at Half-Maximum (FWHM), and Blueshifts are from emission-line profile fitting done in \citet{Hamann+17}. [\oiii] \lam5007 emission blueshift measurements are from \citet{Perrotta+19}, when available, and blueshifted from the reference systemic. \\ \textit{Notes.} $^a$ ERQ's BOSS spectra show strong \mgii\ or \oi\ emission as well as having a CO or \lya\ redshift indicator. $^b$ [\oiii] v$_{98}$ blueshift determined from Keck-OSIRIS observations by Lau~et~al. (in prep.).}
\label{tab:tab_catalogue}
\begin{tabular}{lcccccrrrr}
\hline
    ERQ Name & $z_{em}$ & W3 & \imw & $z_{\text{best}}$ & $z_{\rm best}$ & \civ\ REW & \civ\ FWHM & \civ\ shift & [\oiii] v$_{98}$ \\
    & & (mag) & (mag) & & indicator & (\AA) & (\kms) & (\kms) & (\kms) \\
\hline
J000610.67+121501.2 & 2.3099 & 14.09 & 8.01 & 2.3183 & CO & 107 & 4540 & $-$2260 & $-$5959 \\
J002400.67$-$081110.2 & 2.0633 & 16.19 & 4.63 & 2.0638 & \mgii & 53 & 2042 & 108 & - \\
J005233.24$-$055653.5 & 2.3542 & 16.00 & 6.37 & 2.3631 & CO & 188 & 2451 & $-$1469 & - \\
J011601.43$-$050503.9 & 3.1825 & 15.53 & 6.24 & 3.1875 & \lya-Spike & 94 & 2291 & $-$1240 & - \\
J015222.58+323152.7 & 2.7859 & 15.76 & 5.39 & 2.7925 & \lya-Spike & 136 & 3677 & $-$980 & - \\
J022052.11+013711.1 & 3.1376 & 15.75 & 6.24 & 3.1375 & \lya-Halo & 328 & 2613 & $-$576 & - \\
J080547.66+454159.0 & 2.3258 & 15.51 & 6.32 & 2.3127 & \lya-Spike & 109 & 2667 & $-$524 & $-$4982 \\
J082224.01+583932.8 & 2.5469 & 15.37 & 4.81 & 2.5667 & \lya-Spike & 65 & 5474 & $-$3246 & - \\
J082536.31+200040.3 & 2.0938 & 16.77 & 4.74 & 2.0881 & \lya-Spike & 211 & 3265 & $-$473 & - \\
J082653.42+054247.3 & 2.5734 & 15.18 & 6.01 & 2.5780 & CO & 205 & 2434 & $-$369 & $-$3420 \\
J083200.20+161500.3 & 2.4472 & 14.98 & 6.74 & 2.4249 & CO & 300 & 3082 & $-$297 & $-$5258 \\
J083448.48+015921.1 & 2.5942 & 14.86 & 5.99 & 2.5850 & CO & 209 & 2863 & 198 & $-$4426 \\
J084447.66+462338.7 & 2.2168 & 15.15 & 5.96 & 2.2226 & \lya-Spike & 161 & 1656 & $-$225 & - \\
J085039.50+515831.0 & 1.8914 & 15.62 & 4.97 & 1.9003 & \mgii & 65 & 1145 & $-$1062 & - \\
J085229.65+524730.8 & 2.2674 & 16.48 & 4.74 & 2.2526 & \lya-Spike & 64 & 1291 & $-$322 & - \\
J090014.07+532148.7 & 2.1098 & 17.00 & 7.31 & 2.1042 & \mgii & 47 & 4296 & $-$756 & - \\
J090306.18+234909.8 & 2.2635 & 16.91 & 5.02 & 2.2686 & \mgii & 144 & 2481 & $-$372 & - \\
J091303.90+234435.2 & 2.4195 & 16.41 & 5.31 & 2.4335 & \lya-Spike & 145 & 2190 & $-$448 & $-$2099 \\
J092049.59+282200.9 & 2.2959 & 15.95 & 4.83 & 2.2976 & \lya-Spike & 197 & 1048 & $-$99 & - \\
J092604.08+524652.9 & 2.3467 & 17.13 & 4.71 & 2.3516 & \lya-Spike & 86 & 3053 & $-$634 & - \\
J093638.41+101930.3 & 2.4531 & 15.42 & 6.17 & 2.4523 & \lya-Spike & 172 & 1271 & 196 & - \\
J095033.51+211729.1 & 2.7447 & 16.33 & 5.52 & 2.7430 & \lya-Spike & 272 & 1387 & 86 & - \\
J101326.23+611219.9 & 3.7028 & 15.17 & 5.91 & 3.7061 & \lya-Spike & 281 & 5133 & $-$2945 & - \\
J101533.65+631752.6 & 2.2255 & 16.39 & 5.48 & 2.2337 & \lya-Spike & 130 & 2012 & $-$394 & - \\
J102353.44+580004.9 & 2.5972 & 15.97 & 5.11 & 2.5996 & \lya-Spike & 116 & 2107 & $-$120 & - \\
J104718.35+484433.8 & 2.2751 & 15.57 & 5.30 & 2.2767 & \lya-Spike & 158 & 2521 & $-$33 & - \\
J104754.58+621300.5 & 2.5361 & 16.25 & 5.08 & 2.5566 & \lya-Spike & 105 & 5081 & $-$2382 & - \\
J110202.68$-$000752.7 & 2.6261 & 17.05 & 4.87 & 2.6261 & \lya-Spike & 121 & 3767 & $-$311 & - \\
J111346.10+185451.9 & 2.5160 & 17.07 & 4.62 & 2.5188 & \lya-Spike & 127 & 986 & $-$125 & - \\
J111516.33+194950.4 & 2.7924 & 17.04 & 4.96 & 2.7989 & \lya-Spike & 247 & 1739 & $-$197 & - \\
J111729.56+462331.2 & 2.1317 & 15.55 & 6.26 & 2.1309 & \lya-Spike & 395 & 3053 & $-$390 & - \\
J112124.55+570529.6 & 2.3834 & 14.98 & 5.07 & 2.3885 & \lya-Spike & 28 & 1780 & $-$392 & - \\
J113721.46+142728.8$^a$ & 2.3008 & 15.13 & 4.87 & 2.3025 & CO & 98 & 4734 & $-$2158 & - \\
J113834.68+473250.0$^a$ & 2.3105 & 15.85 & 6.09 & 2.3146 & \lya-Spike & 177 & 3296 & $-$1202 & $-$3928 \\
J113931.09+460614.3 & 1.8202 & 15.30 & 6.44 & 1.8182 & \mgii & 47 & 1239 & 110 & - \\
J114508.00+574258.6 & 2.7904 & 14.27 & 4.84 & 2.8747 & \lya-Halo & 38 & 9103 & $-$8655 & - \\
J121253.47+595801.2 & 2.5841 & 15.85 & 4.95 & 2.5619 & \lya-Spike & 107 & 1402 & 181 & - \\
J121704.70+023417.1 & 2.4163 & 15.43 & 5.59 & 2.4280 & CO & 181 & 2604 & $-$850 & $-$2640 \\
J122000.68+064045.3 & 2.7963 & 16.57 & 4.88 & 2.7732 & \lya-Spike & 113 & 1047 & 63 & - \\
J123241.73+091209.3 & 2.3814 & 14.39 & 6.76 & 2.4050 & CO & 225 & 4787 & $-$3526 & $-$7026 \\
J124106.97+295220.8 & 2.7935 & 16.49 & 5.35 & 2.7976 & \lya-Spike & 138 & 2600 & $-$1382 & - \\
J124738.40+501517.7 & 2.3858 & 16.53 & 4.97 & 2.4014 & \lya-Spike & 135 & 3268 & $-$909 & - \\
J125019.46+630638.6 & 2.4016 & 16.42 & 5.47 & 2.4049 & \lya-Spike & 242 & 1881 & 0 & - \\
J125944.55+240708.3 & 2.1660 & 16.04 & 4.60 & 2.1665 & \mgii & 46 & 4284 & $-$115 & - \\
J130114.46+131207.4 & 2.7867 & 16.26 & 5.11 & 2.7892 & \lya-Spike & 186 & 1877 & $-$347 & - \\
J130630.66+584734.7 & 2.2970 & 16.71 & 5.01 & 2.2986 & \lya-Spike & 331 & 1133 & $-$30 & - \\
J130936.14+560111.3 & 2.5687 & 15.47 & 6.45 & 2.5794 & \lya-Spike & 161 & 3630 & $-$519 & - \\
J131047.78+322518.3 & 3.0168 & 15.05 & 5.26 & 3.0164 & \lya-Spike & 226 & 2794 & $-$914 & - \\
J131330.67+625957.2 & 2.3681 & 17.50 & 4.67 & 2.3714 & \lya-Spike & 82 & 1589 & 49 & - \\
J131351.23+345405.3 & 1.9718 & 15.17 & 4.64 & 1.9706 & \mgii & 18 & 2397 & $-$98 & - \\
J131628.32+045316.2 & 2.1446 & 15.55 & 5.74 & 2.1598 & CO & 63 & 3010 & $-$898 & - \\
J131833.76+261746.9 & 2.2721 & 16.39 & 4.91 & 2.2746 & \lya-Spike & 150 & 1280 & $-$23 & - \\
J132654.95$-$000530.1 & 3.3241 & 15.17 & 4.75 & 3.3068 & \lya-Spike & 77 & 1607 & $-$56 & - \\
\hline
\end{tabular}
\end{center}
\end{table*}
\setcounter{table}{0}
\begin{table*}
\begin{center}
\caption{ERQ Sample Properties (Continued)}
\begin{tabular}{lcccccrrrr}
    \hline
    ERQ Name & $z_{em}$ & W3 & \imw & $z_{\text{best}}$ & $z_{\rm best}$ & \civ\ REW & \civ\ FWHM & \civ\ shift & [\oiii] v$_{98}$ \\
    & & (mag) & (mag) & & indicator & (\AA) & (\kms) & (\kms) & (\kms) \\
    \hline
J133611.79+404522.8 & 2.0793 & 14.10 & 6.51 & 2.0882 & \mgii & 20 & 6154 & $-$2724 & - \\
J134254.45+093059.3$^a$ & 2.3430 & 15.92 & 4.90 & 2.3470 & CO & 66 & 3246 & $-$1558 & $-$4779 \\
J134417.34+445459.4 & 3.0359 & 15.54 & 6.78 & 3.0408 & \lya-Spike & 310 & 2871 & $-$1152 & - \\
J134800.13$-$025006.4 & 2.2495 & 15.91 & 5.70 & 2.2374 & CO & 87 & 3654 & $-$1059 & $-$4221 \\
J135557.60+144733.1$^a$ & 2.7037 & 15.71 & 4.71 & 2.6880 & \lya-Spike & 118 & 2958 & $-$252 & - \\
J135608.32+073017.2 & 2.2691 & 16.52 & 5.08 & 2.2716 & \lya-Spike & 110 & 2043 & 117 & $-$1767 \\
J143159.76+173032.6 & 2.3765 & 16.02 & 5.82 & 2.3880 & \lya-Spike & 177 & 2084 & $-$170 & - \\
J143853.61+371035.3 & 2.3931 & 15.81 & 5.50 & 2.3981 & \mgii & 29 & 3115 & $-$137 & - \\
J144932.66+235437.2 & 2.3428 & 16.79 & 4.73 & 2.3453 & \lya-Spike & 98 & 1352 & 95 & - \\
J145113.61+013234.1 & 2.7734 & 14.77 & 5.67 & 2.8130 & \lya-Halo & 87 & 6231 & $-$3376 & - \\
J145148.01+233845.4 & 2.6214 & 15.01 & 5.51 & 2.6348 & \lya-Halo & 89 & 4166 & $-$3040 & - \\
J150117.07+231730.9 & 3.0254 & 16.20 & 5.90 & 3.0204 & \lya-Spike & 231 & 4035 & $-$1940 & - \\
J152941.01+464517.6 & 2.4201 & 15.92 & 4.82 & 2.4189 & \lya-Spike & 159 & 1896 & 151 & - \\
J153108.10+213725.1 & 2.5689 & 16.92 & 5.23 & 2.5639 & \lya-Spike & 213 & 2767 & $-$253 & - \\
J153446.26+515933.8 & 2.2650 & 17.01 & 4.73 & 2.2658 & \lya-Spike & 127 & 1156 & 95 & - \\
J154243.87+102001.5 & 3.2150 & 15.62 & 6.58 & 3.2166 & \lya-Spike & 114 & 3901 & $-$2303 & - \\
J154743.78+615431.1 & 2.8682 & 16.81 & 4.89 & 2.8674 & \lya-Spike & 128 & 1177 & 10 & - \\
J154831.92+311951.4 & 2.7364 & 16.91 & 4.82 & 2.7462 & \lya-Spike & 127 & 3050 & $-$1151 & - \\
J155725.27+260252.7 & 2.8201 & 16.91 & 4.92 & 2.8217 & \lya-Spike & 56 & 1182 & $-$1 & - \\
J164725.72+522948.6 & 2.7193 & 16.40 & 5.23 & 2.7206 & \lya-Spike & 124 & 1905 & $-$29 & - \\
J165202.64+172852.3 & 2.9425 & 14.91 & 5.39 & 2.9548 & \lya-Halo & 125 & 2403 & $-$1285 & $-$2534 \\
J170558.64+273624.7 & 2.4483 & 15.48 & 5.13 & 2.4461 & \lya-Halo & 157 & 1301 & 267 & - \\
J171420.38+414815.7 & 2.3419 & 16.65 & 4.70 & 2.3303 & \lya-Spike & 130 & 3816 & $-$816 & - \\
J211329.61+001841.7 & 1.9961 & 16.08 & 7.05 & 1.9998 & \lya-Spike & 171 & 1565 & $-$226 & - \\
J220337.79+121955.3 & 2.6229 & 15.50 & 6.21 & 2.6229 & \lya-Spike & 266 & 1070 & 110 & - \\
J221524.00$-$005643.8 & 2.5086 & 16.06 & 6.17 & 2.5093 & CO & 153 & 4280 & $-$1394 & $-$3877 \\
J222307.12+085701.7$^a$ & 2.2890 & 15.65 & 5.60 & 2.2902 & CO & 77 & 3661 & $-$1130 & $-$7760$^b$ \\
J223754.52+065026.6 & 2.6088 & 16.17 & 5.79 & 2.6117 & \lya-Spike & 141 & 1391 & $-$11 & - \\
J232326.17$-$010033.1 & 2.3561 & 15.22 & 7.19 & 2.3805 & CO & 256 & 3989 & $-$2756 & $-$6458 \\
    \hline
\end{tabular}
\end{center}
\end{table*}

\subsection{Systemic Redshift Priorities}

Some ERQs in our sample have available more than one of the systemic redshift indicators mentioned above. We choose a single value of $z_{\rm best}$ for each quasar based on the following priorities. The CO(4-3) emission line has highest priority because it forms (primarily) in dense molecular clouds inside the host galaxies. Although there can be gas motions within the galaxies, we expect the CO line centroids to have velocities similar to the central black holes/quasars (see Hamann~et~al. 2023 in prep. for discussion). Thus we adopt the CO line centroids provided by Hamann~et~al. (2023, in prep.), measured from spatially-integrated spectra, for $z_{\rm best}$ when available. 

Second priority for $z_{\rm best}$, if a CO(4-3) measurement is not available, is a narrow \lya\ emission line arising from the spatially-resolved ERQ halo/circumgalactic medium \citep[as measured from Keck-KCWI observations by][]{Lau+22, Gillette+23a}. Redshifts determined this way are denoted by ``\lya-halo'' in the $z_{\rm best}$ indicator column in Table \ref{tab:tab_catalogue}. We specifically derive redshifts from the spatially-averaged \lya\ emission from extended regions around the ERQs that exclude the central $\sim$1 arcsecond diameter. Excluding the central regions is a precaution to avoid potential i) rapid gas flows near the galactic nucleus, and ii) \lya\ absorption features appearing along direct sight lines toward the central quasar.  

One important result from our studies of ERQ halos with Keck-KCWI \citep{Lau+22,Gillette+23a} was to confirm the speculation by \citet{Hamann+17} that the narrow \lya\ emission spikes seen (with surprising frequency) in aperture spectra of ERQs form in their inner halos. In particular, the narrow \lya\ spikes in aperture spectra consistently have profiles and redshifts very similar to the spatially-resolved halo emission in the KCWI data. 

Our third priority for $z_{\rm best}$ values, when neither CO nor spatially-resolved \lya\ data are available, are a narrow \lya\ spike in aperture spectra of the ERQs. \cite{Lau+22} discussed ERQ J000610$+$121501 is a good example, where a distinct narrow \lya\ spike in aperture spectra clearly forms in the inner halo, and confirmed by mapping data from Keck-KCWI. \cite{Gillette+23a} discusses a wider range of cases, and include some where aperture spectra show the profile of narrow \lya\ spike emitted from the halo blends smoothly into the broad \lya\ emission line of the quasar. We selected the 59 ERQs from \cite{Hamann+17} with \lya\ spikes attributable to halo emission based on the appearance of a single narrow emission peak with FWHM~<~1,000 \kms\ that is well-measured above the noise and shows no indications of overlying \lya\ absorption that might distort the emission spike profile. The centroids of these spikes provide $z_{\rm best}$. Although we give these $z_{\rm best}$ values a lower priority than estimates from spatially-resolved \lya\ halo data, their typical uncertainties are $\leq \sim$200~\kms based on both the line measurement accuracies and our experience comparing these \lya\ spikes to spatially-resolved \lya\ emission from ERQ halos \citep{Lau+22, Gillette+23a}. Quasars with $z_{\text{best}}$ determined this way are denoted by ``\lya-spike'' in Table \ref{tab:tab_catalogue}. 

Lowest priority are the eight ERQs with $z_{\text{best}}$ measurements from their low-ionization \mgii\ \lam 2800 broad emission lines (from a large general study of quasar emission-line blueshifts by Gillette~et~al. 2023c in prep.). We consider these to be the least reliable redshifts because, unlike normal blue quasars, the low-ionization emission lines in ERQs can also be involved fast outflows leading to large blueshifts \citep[see][also Section \ref{sec:sec_discuss} below]{Hamann+19}. None of these quasars with well-measured \mgii\ lines also have CO or \lya\ measures. We note five quasars with CO, or \lya, redshifts and also low-ionization lines with have visibly obvious peak/centroids in \mgii\ or \oi\ \lam1304, indicated in Table \ref{tab:tab_catalogue}.

\subsection{Potential Biases}
\label{sec:sec_biases}

Before discussing emission blueshifts, we consider potential biases in the properties of our ERQ subsamples. The ALMA CO and KCWI \lya -halo observations both tended to favor ERQs with large \civ\ REWs ($\gtrsim$100 \AA) and relatively broad \civ\ profiles (FWHM~$\gtrsim$~2000 \kms ). This represents the majority of ERQs having exotic spectral line/outflow properties \citep[Section 1 of][]{Hamann+17,Perrotta+19}. We also tried to include sources with the reddest \imw\ colors among ERQs. Altogether, however, these samples span a wide range of ERQs because we wanted to examine a range, and because observational constraints tended to randomize the samples with respect to \civ\ properties (e.g., source brightness limits, scheduling constraints, and requirements for low declinations and a very narrow redshift range (near $\sim$2.4) for the ALMA CO observations). Therefore, while these samples do intentionally include some of the most extreme ERQs, e.g., J123241+091209 and J232326$-$010033 with fast [\oiii] outflows \citep{Zakamska+16, Perrotta+19} and J114508+574258 with a previously-known large \civ\ blueshift \citep{Hamann+17}, they should be overall approximately representative of the majority of ERQs with FWHM(\civ)~$\gtrsim$~2000 \kms. We also note that any remaining tendency for large \civ\ REWs in these sample might, if anything, favor small \civ\ blueshifts because large \civ\ REWs are known to correlate strongly with small blueshifts in the general quasar population \citep[][Gillette~et al. 2023c in prep., also Section \ref{sec:sec_outflows_c4}]{Richards+11,Coatman+16,Coatman+19,Rankine+20,Temple+23}. 

\begin{figure}
\includegraphics[width=\columnwidth]{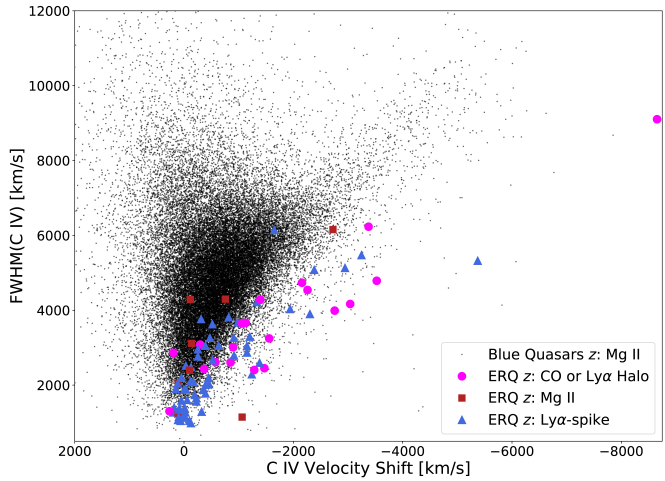}
\vspace{-12pt}
\caption{\civ\ FWHM vs velocity shift of ERQs with the best systemic redshift estimations and blue quasars. Pink dots indicates ERQs with systemic redshift estimates from either CO from ALMA or \lya\ Halo from KCWI, red squares for ERQ redshifts estimates with \mgii\ from BOSS, and blue triangles for ERQ redshifts estimates from narrow \lya\ from BOSS. }
\label{fig:fig_FWHM_shift}
\end{figure}

There is, however, a substantial bias in the sample of 59 ERQs selected to have a narrow \lya\ spike in their BOSS spectra. There are two reasons for this. First, selecting for a narrow spike in \lya\ strongly favors ERQs whose entire \lya\ profile is narrow and whose other ``broad'' emission lines, including \civ\ \lam 1549, are also unusually narrow (e.g., compared to the average for ERQs). Second, narrow \civ\ emission lines correlate strongly with smaller \civ\ blueshifts \citep[Figure \ref{fig:fig_FWHM_shift}, also][and Gillette~et~al. 2023c in prep.]{Richards+11,Coatman+16,Coatman+19,Rankine+20,Temple+23}. The net effect is that this sample is biased towards both narrower \civ\ lines and smaller \civ\ blueshifts than the general ERQ population (see Section \ref{sec:sec_biases} for more discussion).

\section{Emission-line Blueshifts \& Outflow Speeds}
\label{sec:sec_emission_speeds}

\subsection{\civ\ Blueshifts}
\label{sec:sec_outflows_c4}

Table \ref{tab:tab_catalogue} lists \civ\ blueshifts relative to $z_{\rm best}$ for every ERQ in our sample. We use the \civ\ emission-line wavelengths preferred by \cite{Hamann+17}, namely, the midpoint in the fitted line profiles at their half-maximum heights, and a rest wavelength equal to the average for the doublet, 1549\AA. Midpoint wavelengths are the same as the centroid for symmetric profiles, but they avoid possible asymmetries, blueshifted absorption or noise problems in the line wings that would affect the centroid values. 

\begin{figure}
\includegraphics[width=\columnwidth]{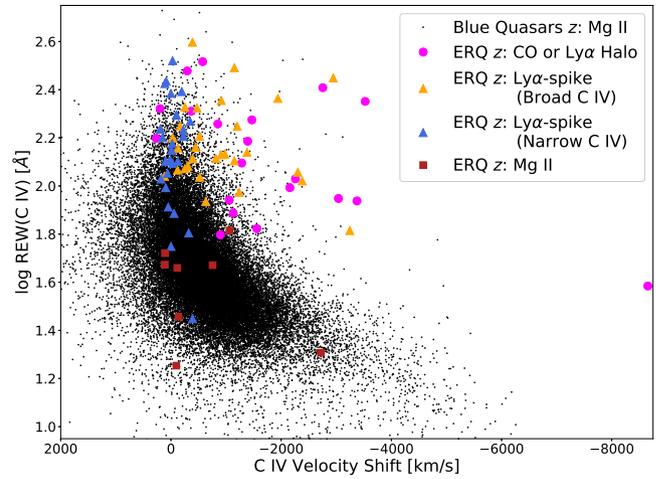}
\vspace{-12pt}
\caption{\civ\ REW vs velocity shift of ERQs with the best systemic redshift estimations and blue quasars. We use the same notation as Figure~\ref{fig:fig_FWHM_shift}, except we separate the group with \lya-spike measurements in to a broad-\civ\ (FWHM $\ge$2000 \kms) and narrow-\civ\ (FWHM <2000 \kms). Broad-\civ\ ERQs can show strongly blueshifted emission, while the narrow-\civ\ ERQs are typically more weakly blueshifted.}
\label{fig:fig_REW_shift}
\end{figure}

Figures \ref{fig:fig_FWHM_shift} and \ref{fig:fig_REW_shift} plot the \civ\ blueshifts versus FWHM(\civ) and REW(\civ), respectively, for ERQs compared to the normal blue quasar sample (Section \ref{sec:sec_sample}). Notice that the ``\lya-spike" sample of ERQs strongly favors narrow \civ\ emission lines (Figure 1). This is not representative of the ERQs overall in \citep[][see also Section 2]{Hamann+17}. However, if we consider only sources with FWHM(\civ)~$\ge$~2000 \kms\ in the \lya-spike sample (orange triangles in Figure \ref{fig:fig_REW_shift}), the distribution of \civ\ blueshifts in that subsample, and its behavior in the 3-dimensional space that includes FWHM(\civ) and REW(\civ), closely resemble the ERQs in the CO and \lya-halo samples (which all have FWHM(\civ)~$\ge$~2000 \kms). 

\begin{figure}
\includegraphics[width=\columnwidth]{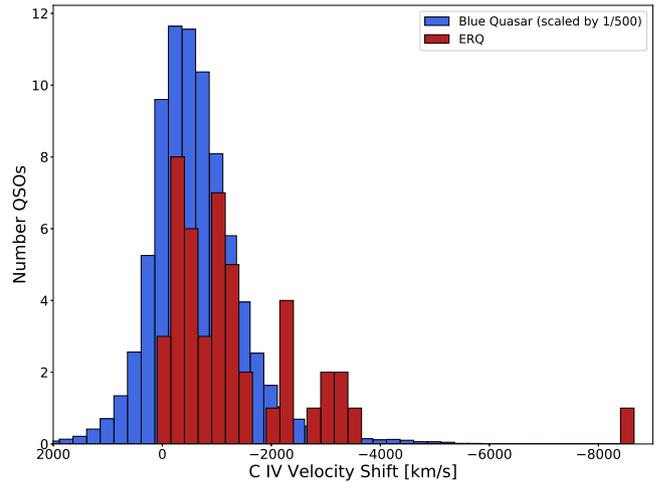}
\vspace{-12pt}
\caption{\civ\ velocity shifts for blue quasars and all ERQs that have reliable systemic redshift estimations from CO or \lya, but excluding ERQs with narrow \civ\ profiles. We scale the much larger blue quasar distribution to compare with our sample. ERQs have a much higher fraction of high velocity \civ\ blueshifts than the normal blue quasar population.}
\label{fig:fig_hist_shift}
\end{figure}

\begin{figure}
\includegraphics[width=\columnwidth]{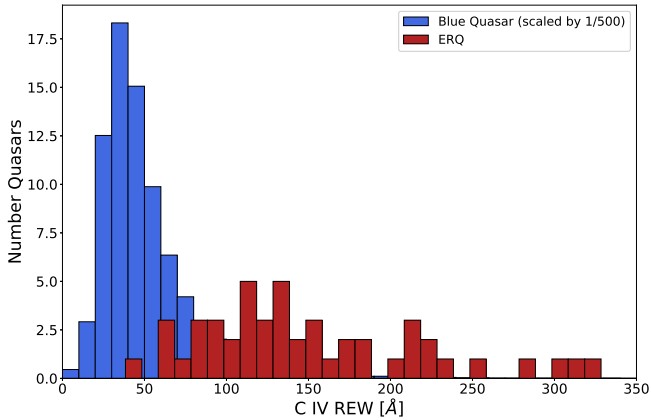}
\vspace{-12pt}
\caption{\civ\ REW for blue quasars and all ERQs that have reliable systemic redshift estimations from CO or \lya, but excluding ERQs with narrow \civ\ profiles. We scale the much larger blue quasar distribution to compare with our sample. ERQs have considerably larger \civ\ REW than the normal blue quasar population.}
\label{fig:fig_hist_rew}
\end{figure}

Figure \ref{fig:fig_hist_shift} and \ref{fig:fig_hist_rew} compare the \civ\ blueshift and REW distributions, respectively, of ERQs to the blue quasar sample. The ERQs in these figures are from the combined sample of 74 sources with $z_{\rm best}$ from either CO, \lya-halo, or a \lya-spike with the additional constraint FWHM(\civ)~$\ge$~2000 \kms.

We can see from these figures that ERQs stand out from normal/blue quasars in several ways. First, ERQs have large \civ\ blueshifts at much larger REW(\civ) and somewhat smaller FWHM(\civ) than normal blue quasars (Figures \ref{fig:fig_REW_shift} and \ref{fig:fig_FWHM_shift}, respectively). It is well known that large \civ\ blueshifts correlate strongly with small \civ\ REWs (Figure \ref{fig:fig_REW_shift}) and large FWHMs (Figure \ref{fig:fig_FWHM_shift}) in normal blue quasars \citep[][Gillette~et~al. 2023c in prep.]{Richards+11,Coatman+16,Coatman+19,Rankine+20,Temple+20,Temple+23}. The ERQs behave differently. They also exhibit a trend for larger blueshifts tied to larger FWHMs (seen in Figure \ref{fig:fig_FWHM_shift}) but, for any given blueshift, the ERQ population is offset toward smaller FWHMs. J114508+574258 is the ERQ with the largest \civ\ blueshift, at $-$8655 \kms\ (Table \ref{tab:tab_catalogue}), and has one of the smallest \civ\ REWs and largest FWHMs among all ERQs \citep{Hamann+17}. These properties are consistent the direction of the trends in blue quasars. However, the \civ\ line in J114508+574258 is substantially broader and at least 4 times larger in REW than expected from trends in blue quasars (Figures \ref{fig:fig_FWHM_shift} and \ref{fig:fig_REW_shift}). The ERQ samples overall are dramatically offset toward larger REWs than blue quasars, especially at large blueshifts. For example, for quasars with \civ\ blueshifts $>$2000 \kms, the average REW among ERQs is 124\AA\ compared to only 24\AA\ for the blue quasar sample. 

Another important difference is the larger fraction of ERQs with large blueshifts compared to the blue quasar population (Figure \ref{fig:fig_hist_shift}). For example, if we consider all ERQs with well-measured $z_{\rm best}$ in our study (excluding only those based on \mgii), the fraction with blueshifts $\ge$2000 \kms\ is 12 out of 82, or 14.63 percent per cent. If we further exclude sources with FWHM(\civ)~<~2000 \kms\ from our \lya -spike sample, to be more representative of the majority of ERQs in \cite{Hamann+17}, we find that 12 out of 54 (22.22 per cent) have blueshifts $>$2000~\kms. In contrast, only 0.23 per cent of the 39909 blue quasars in our comparison sample have blueshifts in this range. \citet{Richards+11} find a similar number, with only 21 (0.13 per cent) out of 15,779 quasars having blueshift $\ge$~2000 \kms\ in their SDSS sample. 

\subsection{Revised [\oiii] Outflow Speeds}
\label{sec:sec_outflows_o3}

Table 1 provides recomputed [\oiii] outflow velocities, v$_{98}$ when available, using our improved systemic redshifts. The [\oiii] data are from \cite{Perrotta+19} except for one quasar, J222307+085701, for which we present an [\oiii] measurement for the first time (from Lau~et~al. in prep.). Systemic redshifts used previously by \cite{Perrotta+19} are based on the best redshift indicator available at that time, namely, either \hb, low-ionization UV emission lines, or distinct narrow components in the [\oiii] (see their Table \ref{tab:tab_catalogue}). Although not discussed explicitly by those authors, a few of the redshifts based on ``low-ions" also considered a narrow core in the \lya\ emission line, if one was present, similar to our analysis in Section \ref{sec:sec_sample}. 

Out of 15 ERQs with [\oiii] data from \cite{Perrotta+19} in our samples, 10 have revised redshifts different from the previous values by $\lesssim$200 \kms. This is generally good agreement for the purpose of studying high-speed quasar outflows. However, 4 have larger revised redshifts by $\gtrsim$500 \kms\ and 3 of those have larger revised redshifts by $\gtrsim$1000 \kms. This implies substantially larger outflow speeds than the previous estimates. In the 4 ERQs with the largest changes, the [\oiii] v$_{98}$ values jumped from $-$2034 to $-$2534 \kms\ in J165202+172852, from $-$5480 to $-$6458 \kms\ in J232326$-$010033, from $-$2872 to $-$3877 \kms\ in J221524$-$005643, and from $-$5580 to $-$7026 \kms\ in J123241+091209 \citep[cf. Table 1 here and Table 4 in][]{Perrotta+19}. All 4 of these ERQs with the largest shifts do not have narrow emission features in the quasar spectra, such that the previous redshift estimates came from \hb\ or low-ionization UV emission lines. 

Five quasars with CO, or \lya, redshifts and low-ionization lines with have visibly obvious peak/centroids in \mgii\ or \oi. We measure velocity shifts of these low-ionization lines with respect to the most reliable emission lines available. These ERQs include J113721+142728, J113834+473250, J134254+093059, J135557+144733, and J222307+085701, with respective relative velocities of $-$218, $-$62, 145, 278, and $-$285\kms. We discuss implications of these velocity shifts in Section \ref{sec:sec_discuss}.

One ERQ in our study with [\oiii] data from \cite{Perrotta+19}, J083448+015921, has a revised redshift that is lower by 653 \kms. This revision changes the v$_{98}$ estimate from $-$5079\kms\ in \citet{Perrotta+19} to $-$4426 \kms\ here. This is surprising because both the CO(4-3) emission line and the low-ionization UV lines in the quasar spectrum, notably \oi\ \lam 1304, appear reasonably strong and well-measured. \cite{Perrotta+19} do not provide uncertainties on their redshift estimates, but a visual inspection suggests that they should be <200 \kms\ (at 3$\sigma$). The formal uncertainty in the CO measurement be Hamann~et~al. (2023 in prep.) is 7.3 \kms. It seems unlikely that the redshift difference is caused by infall in the low-ionization broad emission lines relative to the quasar. We conclude that it might be due to a real kinematic offset between the quasar and the molecular gas emitting CO(4-3) in the quasar's host galaxy. 

\section{Summary \& Discussion}
\label{sec:sec_discuss}

We present a sample of 82 ERQs that have improved estimates for redshift in order to better constrain outflow velocities from comparisons to previous line measurements. These comparisons are subject to selection biases (see Section \ref{sec:sec_biases}), but overall we confirm ERQs have a higher incidence of large \civ\ blueshifts accompanied by large REWs and smaller line widths than blue quasars. Blueshfts >2000 \kms\ are present in 12/54 (22.22 per cent) of ERQs with the most robust $z$ indicators. ERQs with blueshifts >2000 \kms\ are substantially offset in \civ\ REW and FWHM from typical blue quasars in the same velocity range, with ERQ averages of REW = 124\AA\ and FWHM = 5274\kms, compared to blue quasar averages REW = 24\AA\ and FWHM = 6973\kms. 

Our systemic redshifts compared to the previous estimates by \cite{Perrotta+19} already identifies 4 out of 15 ERQs in our sample with blueshifts in their \hb\ and low-ionization UV lines ranging from $-$500 to $-$1500 \kms. This is a lower limit to the true fraction of ERQs with large blueshifts in these lines because some of the estimates in \cite{Perrotta+19} relied on narrow emission spikes in [\oiii], which agree well with our redshift estimates because they also form in extended environments around the quasars. Thus it appears that a significant fraction of ERQs have large blueshifts throughout their broad emission-line regions. This property of ERQs differs from the situation in normal blue quasars, where fast outflows in the broad emission-line regions are both much rarer than ERQs and primarily limited to the high-ionization gas emissions \citep[e.g., \civ,][]{Richards+11,Coatman+19,Rankine+20}. 

The extreme nature of the outflows in ERQs might explain some of their other spectral properties, such as the large \civ\ REWs, e.g., owing to more extended broad emission-line regions that have larger covering factors and reprocess more continuum light from the central quasars, and their peculiar wingless \civ\ profiles, e.g., if they form mostly in outflows instead of virialized gas near the inner accretion disk \citep{Hamann+17}. It is also interesting to consider that the exceptionally fast [\oiii] winds in ERQs might be an outer, lower-density extension to the broad emission-line outflows. 

It remains unclear why ERQs tend to have faster/more powerful outflows than normal blue quasars, but there are two factors that might contribute. One is higher Eddington ratios, which can provide a greater radiative driving force compared to gravity. Another is softer UV continuum, where the weaker far-UV flux helps to maintain moderate ionization levels and substantial opacities in the outflow for radiative driving to be seen in the near-UV. There is strong observational evidence for both of these factors leading to faster outflows in normal blue quasars. For example, the \heii\ \lam 1640 emission-line REW roughly measures the ionizing flux at energies $h\nu > 54$ eV relative to the near-UV continuum, on which the line sits, inversely correlates with larger \civ\ emission-line blueshifts \citep[][Gillette~et~al. 2023c in prep.]{Richards+11,Rankine+20,Temple+23} and faster \civ\ broad absorption-line (BAL) outflows \citep{Baskin+13,Hamann+19,Rankine+20}. This inverse relation is consistent with softer UV spectra playing an important role in outflows, and is indicated in hyper-luminous blue quasars \citep{Vietri+18}. It has been shown that larger Eddington ratios can also correlate with larger \civ\ blueshifts \citep[][Gillette et al. 2023c in prep.]{Baskin+05,Coatman+16,Rankine+20,Temple+23}. 

More work is needed to determine if these trends also apply to ERQs and, moreover, if their extreme outflows result from them being at an extreme end of the trends found in blue quasars (e.g., with larger Eddington ratios or softer UV continua). Previous studies have found no significant trend in [\oiii] outflow speed 
with Eddington ratio (as measured from \hb) among ERQs \citep[see Figure 7 in][]{Perrotta+19}. Unfortunately, there are unique obstacles to testing these trends for ERQs. One is that their emission-line REWs are anomalously large, and \heii\ is at least partly involved in that tendency \citep[see Figure 8 in][]{Hamann+17}, which could confuse the relationship of this line with larger \civ\ blueshifts. Another challenge is that Eddington ratios require black hole mass estimates, which are derived normally from one of the broad emission lines like \mgii. We have shown above that ERQ lines like \mgii\ could have kinematics dominated by outflows, instead of virial motions in the local gravity, and thus making them unreliable for black hole mass determinations. Furthermore, obtaining bolometric luminosities necessary for Eddington ratios poses a greater challenge for ERQs compared to normal blue quasars, because their intrinsic SEDs are potentially atypical. Nonetheless, we confirm the extreme properties of ERQ outflows, which motivates future efforts to understand what is the cause of their extreme nature and potential impacts on the host galaxy. 

\section*{Acknowledgements}

JG, FH, and MWL acknowledge support from the USA National Science Foundation grant AST-1911066. The data presented herein were obtained at the W. M. Keck Observatory, which is operated as a scientific partnership among the California Institute of Technology, the University of California and the National Aeronautics and Space Administration. The Observatory was made possible by the generous financial support of the W. M. Keck Foundation. Data presented herein were partially obtained using the California Institute of Technology Remote Observing Facility. The authors wish to recognize and acknowledge the very significant cultural role and reverence that the summit of Maunakea has always had within the indigenous Hawaiian community.  We are most fortunate to have the opportunity to conduct observations from this mountain.

\section*{Data Availability}

The data are available upon request.


\bibliographystyle{mnras}
\bibliography{ms}


\bsp	
\label{lastpage}
\end{document}